\documentclass[english,aps,preprint,nofootinbib]{revtex4}
\usepackage[T1]{fontenc}
\usepackage[latin1]{inputenc}
\usepackage{color}
\usepackage{graphicx}
\usepackage{amssymb}

\makeatletter

\newcommand{\bee}{\begin{equation}}
\newcommand{\ee}{\end{equation}}
\newcommand{\beea}{\begin{eqnarray}}
\newcommand{\eea}{\end{eqnarray}}

\usepackage{babel}
\makeatother

\begin{document}

\title{Low energy chiral constants from epsilon-regime simulations with
improved Wilson fermions}

\preprint{HU-EP-08/23;SFB/CPP-08-35}

\author{Anna Hasenfratz}

\email{anna@eotvos.colorado.edu}

\affiliation{Department of Physics, University of Colorado, Boulder, Colorado-80309-390,USA}

\author{Roland Hoffmann}

\email{hoffmann@pizero.colorado.edu}

\affiliation{Bergische Universit\"at Wuppertal, Gaussstra{\ss}e~20, 42219 Wuppertal,
Germany}

\author{Stefan Schaefer}

\email{sschaef@physik.hu-berlin.de}

\affiliation{Institut für Physik, Humboldt Universität, Newtonstra{\ss}e~15, 12489
Berlin, Germany}

\begin{abstract}
We present a lattice QCD calculation of the low-energy constants of
the leading order chiral Lagrangian. In these simulations the epsilon
regime is reached by using tree-level improved nHYP Wilson fermions
combined with reweighting in the quark mass. We analyze two point
functions on two ensembles with lattices of size $(1.85{\rm fm})^{4}$
and $(2.8{\rm fm})^{4}$, and at several quark mass values between
4 and 20 MeV. The data are well fitted with next-to-leading order
chiral perturbative formulas and predict $F=90(4)$MeV and $\Sigma^{1/3}=248(6)$MeV
in the $\overline{{\rm MS}}$ scheme at 2 GeV. 
\end{abstract}
\maketitle

\section{Introduction}

At low energies, quantum chromodynamics can be described by a low-energy effective theory, chiral perturbation theory\cite{Gasser:1983yg},
$\chi$PT. To leading order, it has two {\it a priori} unknown parameters,
the chiral condensate and the pion decay constant. In this paper,
we compute these constants by an {\it ab initio} computation. Our strategy
itself is not new: we compute two-point functions in the epsilon regime
and fit them to their $\chi$PT~predictions. However, for chiral
perturbation theory to be an accurate description of low-energy phenomena,
we need calculations on the QCD side at very low pion masses, probably
lower than typically reached in current simulations, and at the same time on large 
enough volumes to control higher order corrections.

Basically, there are two regions in parameter space where calculation
of chiral perturbation theory can be carried out: one is the so-called
$p$-regime where we are essentially at infinite volume. The pion
wave length is much smaller than the size of the box, and the small
corrections due to its finite extent can be taken into account analytically.

The other region is the $\epsilon$-regime\cite{Gasser:1987ah}.
There the pion wave length is much larger than the spatial and temporal
size of the lattice. One can carry out the integrals over the (effectively)
constant pion modes exactly and arrive at a different power counting.
This regime has considerable appeal. First of all, even at next-to-leading order, only the two leading order low-energy constants of
chiral perturbation theory enter into observables\cite{Hansen:1990un,Hansen:1990yg}.
Second, being fundamentally a finite volume regime, once one has
reached it, lowering the pion mass actually improves the validity
of $\chi$PT at a given order, whereas in the $p$-regime one needs
to increase the volume at the same time.

The problem, however, lies in the task to actually reach this regime
in numerical simulations. Because the epsilon expansion is in powers
of $1/(FL)^{2}$, it turns out that we actually need a fairly big
volume and therefore a very small quark mass to satisfy the condition
$m_{\pi}\ll1/L$. This poses many algorithmical problems. In a recent
publication \cite{Hasenfratz:2008fg} we have presented a setup with
which we can actually reach the regime of small quark masses at moderate
cost avoiding many of the problems a more direct approach would face.
We propose to use smeared link improved Wilson fermions \cite{Hasenfratz:2007rf,Schaefer:2007dc}
to generate an ensemble of gauge configurations above the desired
quark mass. From this ensemble, we reweight to the quark mass which
we actually want to reach. Using this method, we manage to reduce
the sea quark mass by roughly a factor of 2 to 4.

Reweighting has another advantage apart from
the actual possibility to go to such light quarks. At very small quark
mass statistical fluctuations grow dramatically because the mass no
longer provides an infrared cutoff. Therefore, very small eigenvalues
of the Dirac operator can appear, which are suppressed by the fermion
determinant, but also lead to large values of the measured observables.
We thus have an anticorrelation between the weight and the function
value which means importance sampling breaks down. Reweighting avoids
this problem. The small eigenvalues are not as efficiently suppressed,
the region of large signal gets over-sampled and the error is actually
reduced. 

One might wonder how the explicit chiral symmetry
breaking of Wilson fermions influence simulations done with very light quarks in the epsilon regime.
Unlike in the $p$-regime, the finite volume of the $\epsilon$-regime provides 
the system  with an effective, dynamical IR cut-off. The chiral
symmetry violations will remain managable even in the chiral limit
as long as they are small compared to the inverse lattice size.
The numerical study of the continuum limit to prove this expectation will
be the subject of a future work.

In the current paper, we apply reweighting to a large volume data
set and provide measurements of the low-energy constants $F$ and
$\Sigma$. We already mentioned that this type of simulation is not
new. A number of quenched studies have been carried out \cite{Bietenholz:2003bj,Giusti:2003iq,Giusti:2004yp,Fukaya:2005yg,Bietenholz:2006fj,Giusti:2008fz}
using overlap fermions, which proved the feasibility of the method
but also highlighted the need for sufficiently large volume. Recently,
there also have been computations with dynamical overlap\cite{Fukaya:2007pn}
and twisted mass\cite{Jansen:2007rx} fermions to which we will compare
our results in the conclusion. For a recent review of these calculations
see Ref.~\cite{Necco:2007pr}. The strength of our calculation is
that we compare two different volumes, of which the larger one has
not been reached in previous studies. The simulations in this work
were fairly inexpensive, and it is within reach to repeat the calculation
at a finer lattice spacing to verify scaling.

The outline of this paper is the following: We first describe in Sec.~\ref{sec:lat}
the set-up of our simulations, the generation of the ensemble of gauge
configurations and the details of our reweighting procedure. The relevant
renormalization constants are computed in Sec.~\ref{sec:ren} using
the RI-MOM scheme. In Sec.~\ref{sec:eps} we collect the $\epsilon$-regime
formulas, the procedure of the extraction of the low-energy constants
and give the results.

\section{Simulations\label{sec:lat}}

The numerical simulations for this project were done with 2 flavors
of nHYP smeared Wilson-clover fermions and one-loop Symanzik improved
gauge action. The action and the simulation method are described in
details in Refs. \cite{Hasenfratz:2007rf,Schaefer:2007dc}. We
use tree-level $c_{SW}=1.0$ clover coefficient, so our action is
not fully $\mathcal{O}(a)$ improved. Based on our quenched investigation~\cite{Hoffmann:2007nm},
we expect that a nonperturbatively improved action would require
$c_{SW}\lesssim1.2$, so even with only tree-level improvement the
$\mathcal{O}(a)$ corrections are likely small. We have generated
two sets of gauge ensembles, both at gauge coupling $\beta=7.2$.
The first set consists of $16^{4}$ configurations at $\kappa=0.1278$,
the second $24^{4}$ configurations at $\kappa=0.12805$. We have
180 and 154 thermalized configurations, separated by 5 trajectories
at the two volumes. The autocorrelation is 3-4 trajectories for the
plaquette, and about the same for the two point functions. Preliminary
results using the first set were already reported in Ref. \cite{Hasenfratz:2008fg}.

We set the lattice scale from the static quark potential, using $r_{0}=0.49$
fm for the Sommer parameter. On both configuration sets we found $r_{0}/a=4.25(2)$
(the error is from the larger volume set where we have a better signal
for the potential), giving $a=0.1153(5)$ fm. With this value the
physical volumes are (1.85 fm)$^{4}$ and (2.77 fm)$^{4}.$ Based
on the PCAC quark mass and the pseudoscalar and axialvector renormalization
factors (see Sec.~\ref{sec:ren}), we estimate the renormalized quark
mass in the $\overline{\rm MS}$ scheme at 2 GeV to be 22 and 8.5MeV, respectively. These values, and some other
details of the simulation, are listed in Table \ref{cap:Basics}.
\begin{table}
\begin{tabular}{|c|c|c|c|c|c|}
\hline 
$\kappa$  & $\kappa_{{\rm rew}}$  & $L$  & $N_{{\rm conf}}$  & $a\, m_{{\rm PCAC}}$  & $m${[}MeV]\tabularnewline
\hline
\hline 
0.1278  & 0.1278  & 16  & 180  & 0.0117(3)  & 22\tabularnewline
\hline 
 & 0.1279  & 16  & 180  & 0.0088(5)  & 16.5\tabularnewline
\hline 
 & 0.1280  & 16  & 180  & 0.0058(7)  & 11\tabularnewline
\hline 
 & 0.12805  & 16  & 180  & 0.0047(8)  & 9\tabularnewline
\hline 
 & 0.1281  & 16  & 180  & 0.0028(11)  & 5\tabularnewline
\hline
\hline 
0.12805  & 0.12805  & 24  & 154  & 0.0044(3)  & 8.5\tabularnewline
\hline 
 & 0.12810  & 24  & 154  & 0.0030(3)  & 5.8\tabularnewline
\hline 
 & 0.128125  & 24  & 154  & 0.0024(3)  & 4.2\tabularnewline
\hline 
 & 0.12815  & 24  & 154  & 0.0019(4)  & 3.8\tabularnewline
\hline
\end{tabular}

\caption{The parameters of the simulation. The first column gives the coupling
$\kappa$ of the dynamical simulation, the second the reweighted coupling
$\kappa_{{\rm rew}}$. The last column is the renormalized quark mass
using $m=m_{{\rm {PCAC}}}Z_{A}/Z_{P}$. \label{cap:Basics}}

\end{table}

\begin{figure}
\includegraphics[scale=0.4]{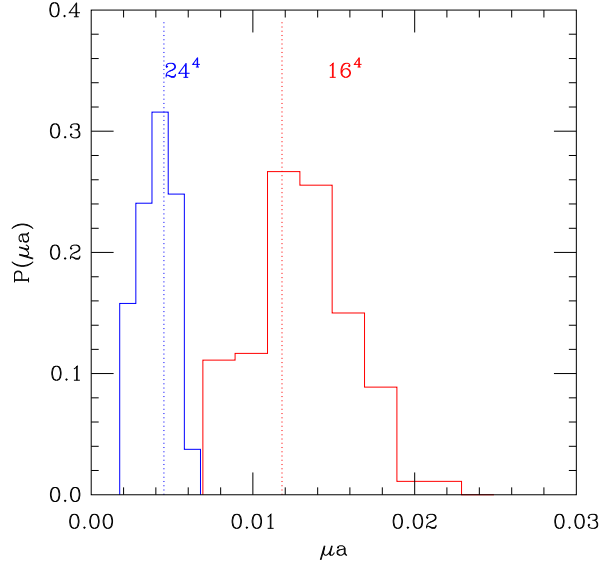}

\caption{The Hermitian gap distribution on the original $16^{4}$ and $24^{4}$
configurations. The dashed lines correspond to $am_{{\rm PCAC}}$
. \label{cap:The-gap-distribution}}

\end{figure}

The dynamical simulations were performed at particularly low quark
masses, even if we consider the relatively large volumes. This is
possible due to the highly improved chiral properties of the nHYP
smeared clover fermions. Figure \ref{cap:The-gap-distribution} shows
the histogram of the absolute value of the lowest Hermitian eigenmode
of the configurations. Simulations with Wilson-like fermions require
a well defined gap between zero and the first eigenmode \cite{DelDebbio:2006cn}.
As the figure shows, our simulations are safe on both volumes, though
very close to the low mass limit. In both cases, the median of the
distribution is about four times its width $\sigma$, which is above
the $3\sigma$ stability criterion of Ref.~\cite{DelDebbio:2006cn}.
However, it also shows that going lower in the quark mass can be very
dangerous because of the algorithm becoming unstable. The same paper
predicts $\sqrt{V}\sigma/a$ to be a scaling quantity. We measure
0.56(4) and 0.77(5) for the $24^{4}$ and $16^{4}$ ensembles respectively.

Starting from the original configurations one can explore a range
of quark masses in fully dynamical systems by reweighting the configurations.
In Ref.~\cite{Hasenfratz:2008fg} we have described an effective
technique to calculate the necessary weight factors. It is a stochastic
calculation, and one must take care not to introduce significant statistical
errors with the stochastic process. We apply three methods, low mode
separation, determinant breakup, and ultraviolet (UV) noise reduction
to control the statistical fluctuations. In both the $16^{4}$ and
$24^{4}$ ensembles we separate 6 low Hermitian eigenmodes. In addition
we break up the determinant to the product of 33 and 60 terms for
each $\Delta\kappa=0.0001$ shift in reweighting on the $16^{4}$
and $24^{4}$ volumes, respectively. To control and remove some of
the UV noise we introduce a pure gauge term in the reweighted action
. This term is just an nHYP plaquette term and has a very small coefficient.
We found that in our system it can be chosen to be proportional to
the shift in $\kappa$, \begin{equation}
\beta_{{\rm nHYP}}=6.0(\kappa-\kappa_{{\rm rew}})\:\label{eq:beta-nhyp}\end{equation}
 on both the $16^{4}$ and $24^{4}$ configuration sets. This value
is so small that there is no difference within the errors in the lattice
spacings or quark masses between $\beta_{{\rm nHYP}}=0$ and Equation \ref{eq:beta-nhyp}.
For further details and the exact definition of the reweighting action
we refer to Ref. \cite{Hasenfratz:2008fg}, especially Equation 17 and
Figure 4.

\begin{figure}

\includegraphics[scale=0.8]{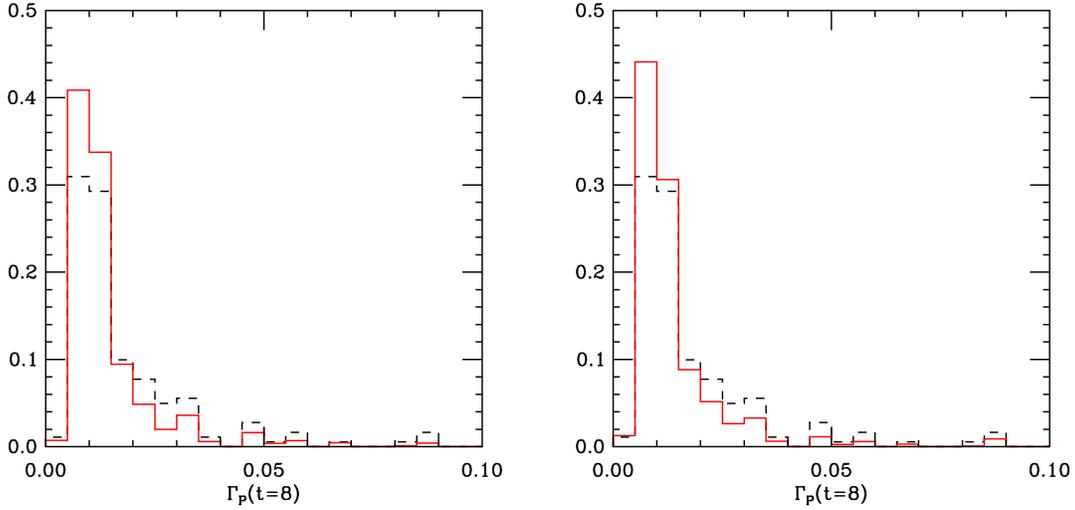}

\caption{The distribution of the pseudoscalar correlator at $t=8$ on the $16^{4}$
ensemble at $\kappa=0.1280$. On both panels the dashed line is the
partially quenched distribution and the solid lines correspond to the
reweighted distributions. Left panel: $\beta_{{\rm nHYP}}=0$, right
panel: $\beta_{{\rm nHYP}}$ as in Equation \ref{eq:beta-nhyp}.\label{fig:overlap}}

\end{figure}
In addition to removing the UV fluctuations, the introduction
of the nHYP plaquette term also increases the overlap between the
original and target ensembles. The largest weight factors are pushed
from the edge of the plaquette distribution to the middle, where the
statistical sampling is better, and the effect is similar for other
observables as well. We illustrate this in Figure \ref{fig:overlap}
with the distribution of the pseudoscalar correlator at $t=8$. All
data correspond to the $16^{4}$ data set at $\kappa=0.1280$. The
left panel shows the reweighted distribution without the nHYP plaquette
term, the right one with $\beta_{{\rm nHYP}}$ as in Equation \ref{eq:beta-nhyp}.
For reference both panels show the partially quenched (unweighted)
distribution. In both cases the overlap between the reweighted and
partially quenched distributions is excellent, there is no sign that
reweighting would prefer region that is poorly sampled by the original
ensemble. The main difference between the reweighted and partially
quenched data is the suppression of the long tail of the latter one,
a quenching artifact. The apparent increase of the reweighted distributions
is mainly due to normalization: the dynamical distribution is narrower,
resulting in a higher peak. 
It is worthwhile to emphasize that including the nHYP plaquette term does not 
introduce any systematic error, rather it improves the overlap between the ensembles, 
especially at larger mass difference.

Even though there is no strong difference between
the two panels of Figure \ref{fig:overlap}, the introduction of
the nHYP plaquette term reduces the statistical errors by up to 40\%
for our lightest $16^{4}$ data set, and the results we present in
the following reflect that. On the other hand, on the $24^{4}$ volumes
within the $\kappa$ range we reweight to there is no difference between
the actions with or without the nHYP plaquette term, and the results
we present here were obtained with $\beta_{{\rm nHYP}}=0$ .

With reweighting it is possible to reach an eigenvalue that is negative
or at least smaller than the typical eigenvalues of the Dirac operator.
We approach that case with our last coupling on the $16^{4}$ ensemble
where at least one configurations has a negative real Dirac eigenvalue
and several has nearly zero eigenvalues, and even more on the lightest
reweighted $24^{4}$ configurations where 5\% of the configurations
have a negative Dirac eigenvalue and even more have a nearly zero
one. %
\begin{figure}
\includegraphics[scale=0.8]{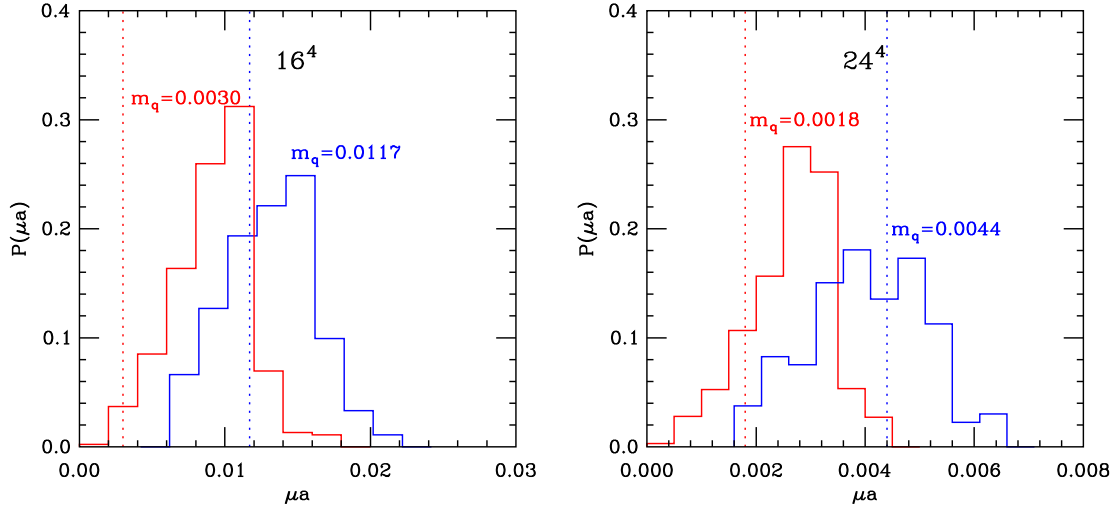}

\caption{The Hermitian gap distribution on the original and lightest reweighted
ensembles for both volumes. The histograms are labeled by the corresponding
lattice quark mass which is also indicated by the dashed lines. \label{fig:gap_distr} }

\end{figure}
These configurations are suppressed by the weight factor, nevertheless
as we will see later one encounters increased statistical errors on
these ensembles. In Figure \ref{fig:gap_distr} we compare the Hermitian
gap distribution on the original and lightest reweighted ensembles
for both volumes. The distribution shifts towards zero but configurations
with negative or near-zero eigenmodes are strongly suppressed, and that
maintains  the  gap. While the PCAC quark mass approches zero, the median of
the gap, controlled by the finite volume,  remains finite.

\section{Renormalization factors\label{sec:ren}}

In order to connect the lattice meson correlators to the physical
ones we have to determine the corresponding renormalization factors.
We used the standard RI-MOM method \cite{Martinelli:1994ty}, where
one calculates bilinear quark operators $\langle p|O_{\Gamma}|p\rangle$
at specific lattice momentum $p^{2}=\mu^{2}$ and matches them to
the corresponding tree-level matrix element. Afterwards the lattice
values are connected to the continuum $\overline{{\rm MS}}$ scheme
perturbatively \cite{Gimenez:1998ue,Chetyrkin:1999pq}. The renormalization
scale $\mu$ has to be much smaller than the lattice cut-off to minimize
lattice artifacts but much larger than the QCD scale for continuum
perturbation theory to work.

Our code to calculate the lattice matrix elements is based on the
one used in Ref. \cite{DeGrand:2005af}. We used 80 propagators from
the $16^{4}$ data set to calculate the vector, axialvector, scalar
and pseudoscalar matching factors in the chiral limit. While most
dynamical calculations do the chiral extrapolations on partial quenched
data \cite{Becirevic:2005ta},\cite{Dimopoulos:2007fn}, we can
do this extrapolation on fully dynamical configurations on our reweighted
ensembles. We used 5 $\kappa$ values, 0.1278, 0.12785, 0.1279, 0.12795
and 0.1280, corresponding to quark masses 10-20 MeV. We extrapolated
the vector, axial and scalar data linearly in the quark mass, though
the data shows no mass dependence within errors. This is not that
surprising, since our quark masses are light. The pseudoscalar density
couples to the Goldstone boson channel and it develops an $\mathcal{O}(1/m)$
singularity in the chiral limit. We subtract this pole assuming a
linear mass dependence for the quantity $m/Z_{P}$. Again, with our
light mass values we expect this assumption to hold, and our data
are indeed consistent with a linear dependence \cite{Becirevic:2004ny,Gattringer:2004iv}.
Nevertheless the subtraction introduces fairly large errors at small
$\mu$.

\begin{figure}
\includegraphics[scale=0.8]{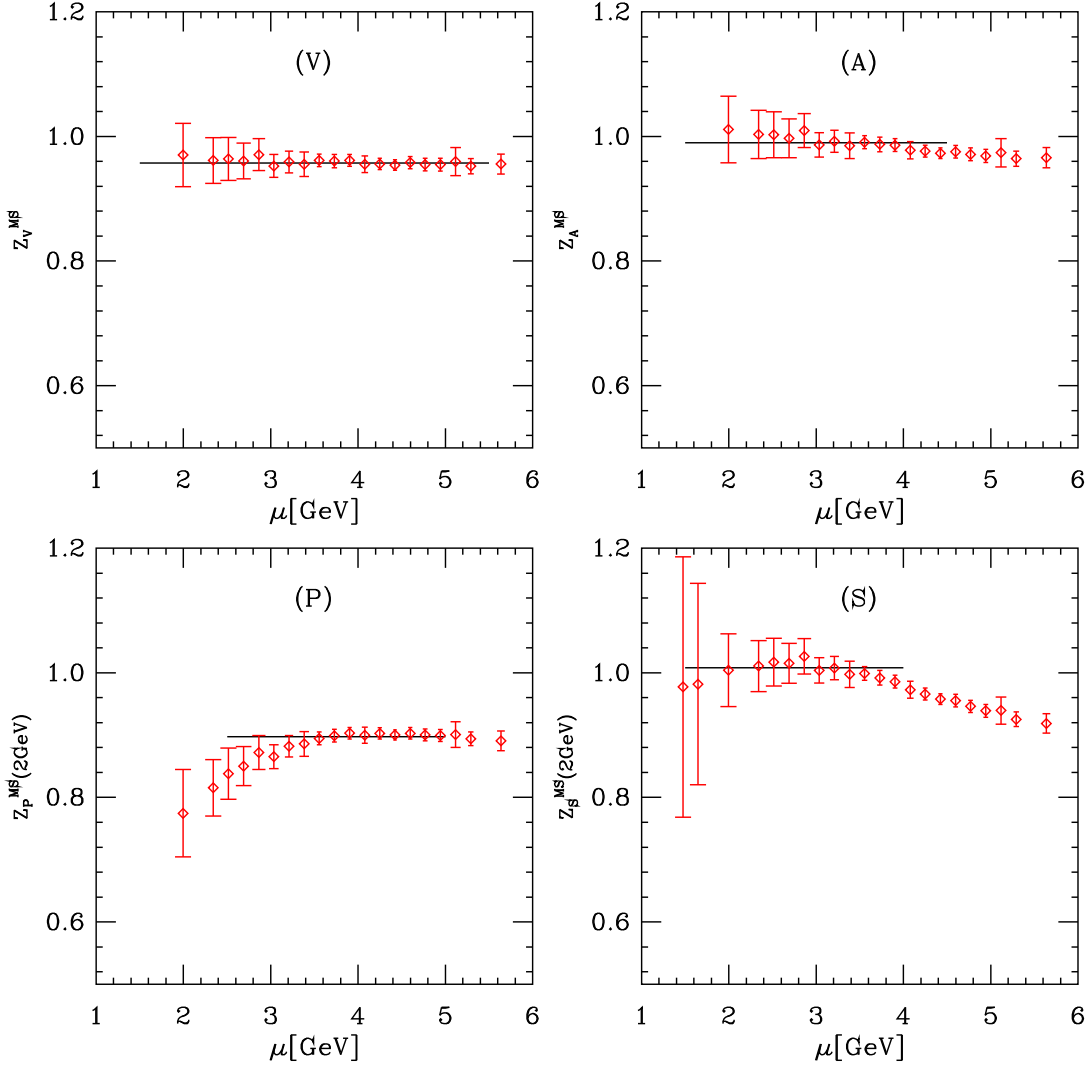}

\caption{The renormalization factors for the vector, axialvector, pseudoscalar
and scalar operators as the function of the lattice energy scale.
All values are converted to the continuum $\overline{{\rm MS}}$ scheme
at $\mu=2$GeV. \label{cap:The-renormalization-factors}}

\end{figure}

In Figure \ref{cap:The-renormalization-factors} we show all four
renormalization factors converted to the $\overline{{\rm MS}}$ scheme
at 2GeV, as the function of the original lattice momentum $p^{2}=\mu^{2}$.
The vector and axialvector factors are scale independent, any deviation
from a constant is due to lattice artifacts. In our case $Z_{V}$
is constant over the whole range, while $Z_{A}$ shows a slight drift
at larger $\mu$ values. Calculations with Wilson and Wilson-like
improved fermions show similar trends for these quantities \cite{Becirevic:2004ny,Becirevic:2005ta,Gattringer:2004iv}.

The scalar and pseudoscalar operators depend on the energy scale.
We connect the lattice data to the continuum one at identical energy,
then, using the known 3-loop expression for the running of the coupling,
run it to $\mu=2$GeV . We plot these values, therefore $Z_{P}$ and
$Z_{S}$ in Figure \ref{cap:The-renormalization-factors} should also
be constant. Because of the subtraction of the Goldstone pole the errors
are large at small $\mu$ for the pseudoscalar, but we find a long,
stable plateau at larger lattice scale values. The scalar operator,
on the other hand, shows quite large lattice artifacts at higher scales.
Again, this trend has been observed before with other actions. The
horizontal lines in Figure \ref{cap:The-renormalization-factors}
indicate the range where we extract the renormalization factors. Our
final values are \begin{eqnarray}
Z_{V}^{\overline{{\rm MS}}} & = & 0.96(1)\nonumber \\
Z_{A}^{\overline{{\rm MS}}} & = & 0.99(2)\nonumber \\
Z_{P}^{\overline{{\rm MS}}}(2{\rm GeV}) & = & 0.90(2)\label{eq:Z_factors}\\
Z_{S}^{\overline{{\rm MS}}}(2{\rm GeV}) & = & 1.01(3)\,.\nonumber \end{eqnarray}
 As a simple check we compare the renormalized quark mass as predicted
form the bare quark mass $m_{}=Z_{S}^{-1}m_{b}$, $1/(2\kappa)-1/(2\kappa_{cr})$,
and from the PCAC mass $m_{r}=Z_{A}Z_{P}^{-1}m_{{\rm PCAC}}$. Fitting
$m_{{\rm PCAC}}$ linearly in $1/(2\kappa)$ we predict $\kappa_{cr}=0.12821$
and from the slope $Z_{P}Z_{S}^{-1}Z_{A}^{-1}=0.94(3)$. This is consistent
from the value obtained from Equation \ref{eq:Z_factors}, 0.90(7). The
fact that all four matching factors are close to one indicates small
perturbative corrections, as it is usually seen with smeared link
actions.

\section{$\epsilon$-regime analysis\label{sec:eps}}

In the $\epsilon$-regime the pion correlation length is large compared
to the linear size of the lattice, the light pseudoscalar mesons
dominate the dynamics. Nevertheless, in order to incorporate the massive
modes the volume has to be large compared to the QCD scale. One assumes
that the quark mass is $m=\mathcal{O}(\epsilon^{4})$ and the inverse
size $1/L=\mathcal{O}(\epsilon)$ ( $L^{4}=V=L_{s}^{3}L_{t}$), but
$1/L\gg\Lambda_{{\rm QCD}}$. The dimensionless quantity $m\Sigma V$,
or equivalently $m_{\pi}^{2}F_{\pi}^{2}V$, is kept order one. Chiral
perturbation theory predictions are organized in power of $\epsilon^{2}$
or $1/(FL)^{2}$. Predictions for various meson correlators are known
to next-to-leading (NLO) order ($\mathcal{O}(\epsilon^{4})$), except
for the pseudoscalar that has been calculated up to $\mathcal{O}(\epsilon^{6})$.
In our fits we use the NLO predictions as those depends only on two
low-energy constants, $\Sigma=\lim_{m\to0}\langle\bar{q}q\rangle$
and $F=\lim_{m\to0}F_{\pi}$\cite{Hasenfratz:1989pk,Hansen:1990un,Hansen:1990yg}.
These $\chi$PT results are based on a chiral (continuum) action.
One expects extra terms, due to the explicit chiral symmetry violation
of the Wilson fermion action, in our situation. However these corrections
typically show up at the same order as the higher order chiral constants
$L_{3},$$L_{4}$, i.e. only at next-to-next-to-leading order in the
epsilon regime.

As the quark mass decreases in a large volume (p-regime)
simulation, the chiral symmetry breaking effects of Wilson fermions
get large compared to the mass, and that can create large lattice
artifacts. In practice the continuum limit has to be taken
before the chiral limit. The situation is different in the $\epsilon$-regime, where
the finite volume of the system creates an infrared cutoff even at
vanishing quark mass. This effect is well illustrated by the Hermitian
gap distribution in Figure \ref{fig:gap_distr}. While in infinite
volume one expects the median of the gap to scale with the mass $\bar{\mu}=Z_{A}m_{{\rm PCAC}}$
\cite{DelDebbio:2005qa}, in the $\epsilon-$ regime $\bar{\mu}$,
governed by the IR cutoff of the volume, remains finite while $m_{{\rm PCAC}}\to0$.
This is clearly the case in our simulations. Therefore one does not need a chiral
action to study the epsilon regime, though the explicit symmetry breaking
effects should be small compared to the inverse lattice size. As long
as the volume is large enough that the NLO relations describe the
two-point functions, continuum $\chi$PT results can be used to analyze
Wilson fermion data. Since separating topological sectors with Wilson
fermions is not always possible, we analyze our data averaged over
the topological charge.

For completeness we give the relevant formulas for two degenerate
flavors, averaged over the topological charge. The isotriplet pseudoscalar meson correlator up to $\mathcal{O}(\epsilon^{4})$ is \begin{eqnarray}
\Gamma_{P}(t) & = & \frac{1}{L_{s}^{3}}\int d^{3}x\langle P(x)P(0)\rangle\nonumber \\
 & = & \Sigma^{2}\Big(a_{p}+\frac{L_{t}}{F^{2}L_{s}^{3}}b_{p}h_{1}(\frac{t}{L_{t}})+\mathcal{O}(\epsilon^{4})\Big)\,,\label{eq:eps-pion-corr}\end{eqnarray}
 where $P(x,t)=\bar{\psi}\frac{1}{2}\lambda^{i}\gamma_{5}\psi$ is
the pseudoscalar density operator and \begin{eqnarray}
a_{p} & = & \frac{\rho}{8}I_{1}(u)\,.\label{eq:chipt_coeff}\\
b_{p} & = & 1-\frac{1}{8}I_{1}(u)\,,\nonumber \end{eqnarray}
 with \[
\rho=1+\frac{3\beta_{1}}{2(FL)^{2}}\]
 the shape factor ($\beta_{1}=0.14046$ for our symmetric geometry
) and \[
u=2\, m\Sigma V\,\rho\,.\]
 $I_{1}$ can be expressed in terms of Bessel functions, $I_{1}(u)=8Y'(u)/(uY(u))$.
It decreases smoothly from 2 at $u=0$ to 0.68 at $u=10$, the largest
value we encounter. The function \begin{equation}
h_{1}(\tau)=\frac{1}{2}\big[(\tau-\frac{1}{2})^{2}-\frac{1}{12}\big]\label{eq:h1}\end{equation}
 describes the quadratic time dependence. The pseudoscalar correlator
is dominated by $\Sigma$, the dependence on $F$ is only through
the $\mathcal{O}(\epsilon^{2})$ term.

The flavor triplet axialvector current correlator at NLO is \begin{eqnarray}
\Gamma_{A}(t) & = & \frac{1}{L_{s}^{3}}\int d^{3}x\langle A_{0}(x)A_{0}(0)\rangle\nonumber \\
 & = & \frac{F^{2}}{V}\Big(a_{a}+\frac{L_{t}}{F^{2}L_{s}^{3}}b_{a}h_{1}\Big(\frac{t}{L_{t}}\Big)+\mathcal{O}(\epsilon^{4})\Big)\,,\label{eq:eps-axial-corr}\end{eqnarray}
 with $A_{0}(x,t)=\bar{\psi}\frac{1}{2}\lambda^{i}\gamma_{0}\gamma_{5}\psi$
and \begin{eqnarray}
a_{a} & = & 1-\frac{1}{4}I_{1}(u)+\frac{\beta_{1}}{(FL)^{2}}\Big(2-\frac{1}{2}I_{1}(u)\Big)-\frac{L_{t}}{F^{2}L_{s}^{3}}\frac{k_{00}}{2}I_{1}(u)\,,\nonumber \\
b_{a} & = & \frac{1}{8}u^{2}I_{1}(u)\,.\label{eq:axial_coeff}\end{eqnarray}
 In our $L_{s}=L_{t}$ case $k_{00}=\beta_{1}/2$. The axialvector
correlator is dominated by $F,$ the dependence on $\Sigma$ enters
only through the combination $m\Sigma V$.

The $\epsilon$-expansion formulas are systematic expansions in the
parameter $1/(FL)^{2}=\mathcal{O}(\epsilon^{2})$, but depend on the
$\mathcal{O}(1)$ quantity $m\Sigma V$. In our simulation we explore
the range $m\Sigma V\approx0.7\,-\,5.0$. Large values introduce large
NLO and NNLO corrections to the correlators, and at some point one
transitions into the large volume $p$-regime. Only by examining the
fit results will we be able to decide what range of $m\Sigma V$ values
are acceptable in the $\epsilon$-regime.

The lattice correlators have to be multiplied by the renormalization
factors $Z_{P}^{2}$ and $Z_{A}^{2}$ to obtain the continuum ones
in Equations \ref{eq:eps-pion-corr} and \ref{eq:eps-axial-corr}, while
in the product $m\Sigma V$ the quark mass can be expressed in terms
of the PCAC mass as $m=m_{{\rm PCAC}}Z_{A}/Z_{P}$. In our fit we
use the combinations \begin{equation}
\Gamma_{A}=Z_{A}^{2}\Gamma_{A}^{({\rm latt})}\label{eq:axial}\end{equation}
 and \begin{equation}
m^{2}\Gamma_{P}=Z_{A}^{2}m_{{\rm {PCAC}}}^{2}\Gamma_{P}^{({\rm latt})}\label{eq:pseudo}\end{equation}
 that depend only on $F$ and $m\Sigma V$, and do a combined fit
to Equations \ref{eq:axial} and \ref{eq:pseudo}. %
\begin{figure}
\includegraphics[scale=0.7]{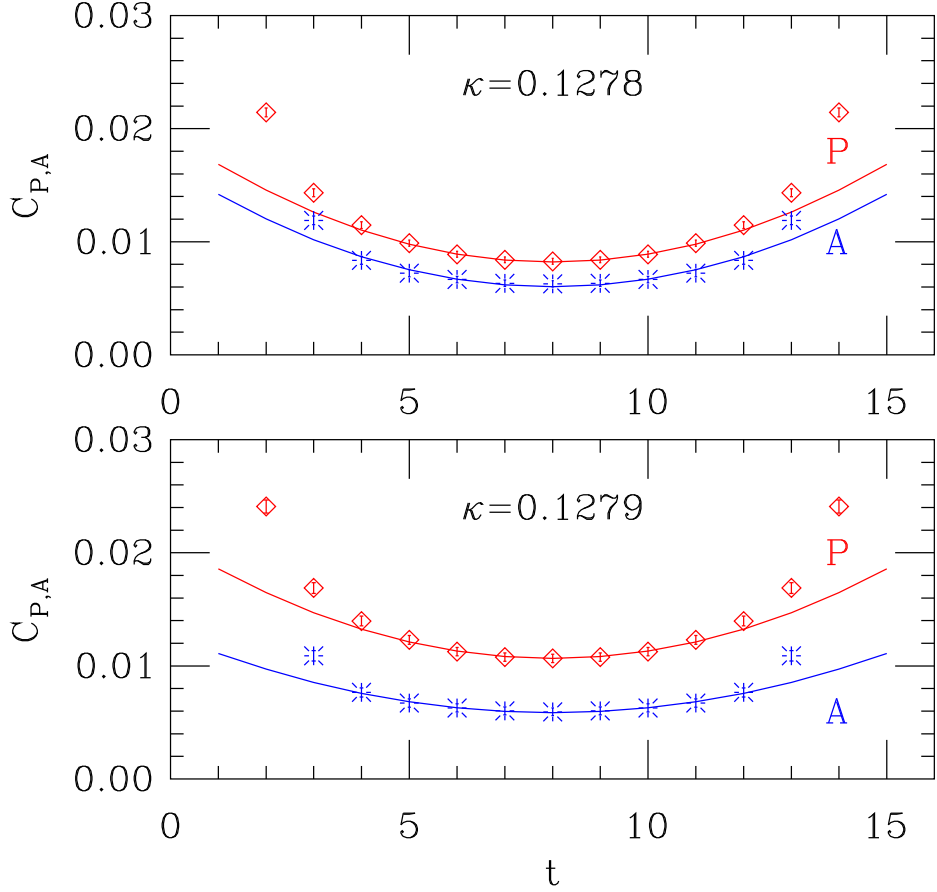}

\caption{The pseudoscalar ( red diamonds) and axialvector (blue bursts) lattice
correlators and the combined fit results at $\kappa=0.1278$ and $0.1279$
on the $16^{4}$ data set. The axialvector correlators are multiplied
by the factor 50 to better match the scale of the pseudoscalar.\label{cap:corr-16-1}}

\end{figure}

\begin{figure}
\includegraphics[scale=0.7]{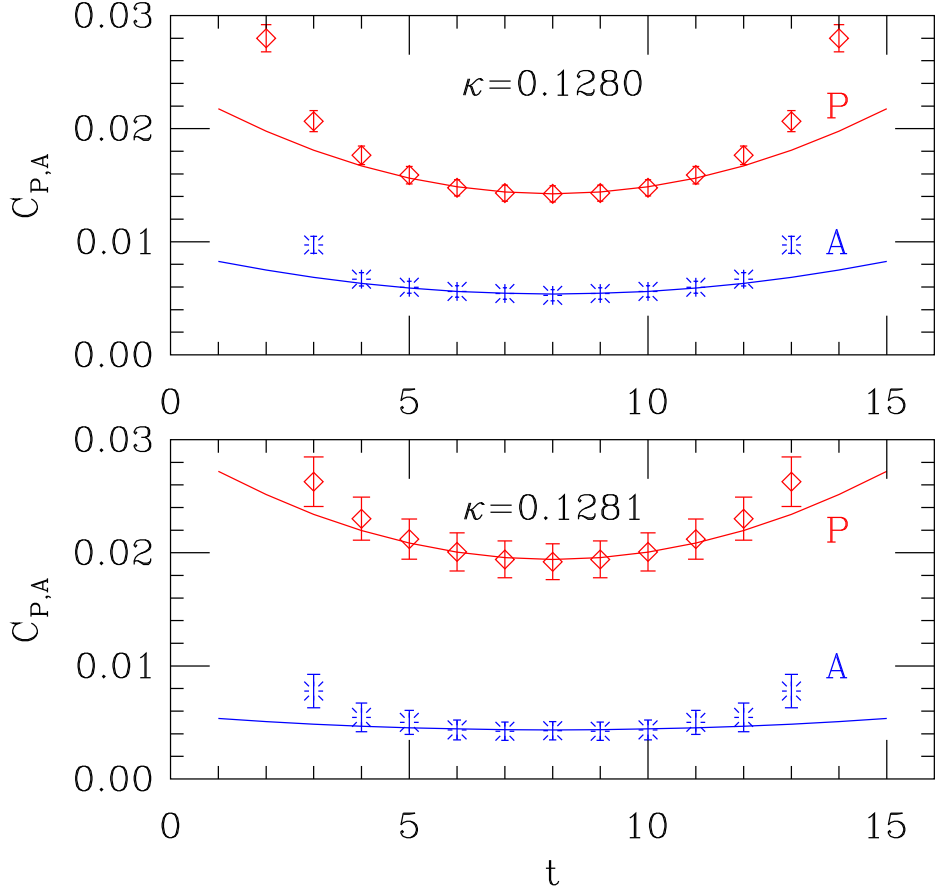}

\caption{Same as Figure \ref{cap:corr-16-1} but at $\kappa=0.1280$ and $0.1281$.\label{cap:corr-16-2}}

\end{figure}

\begin{figure}
\includegraphics[scale=0.7]{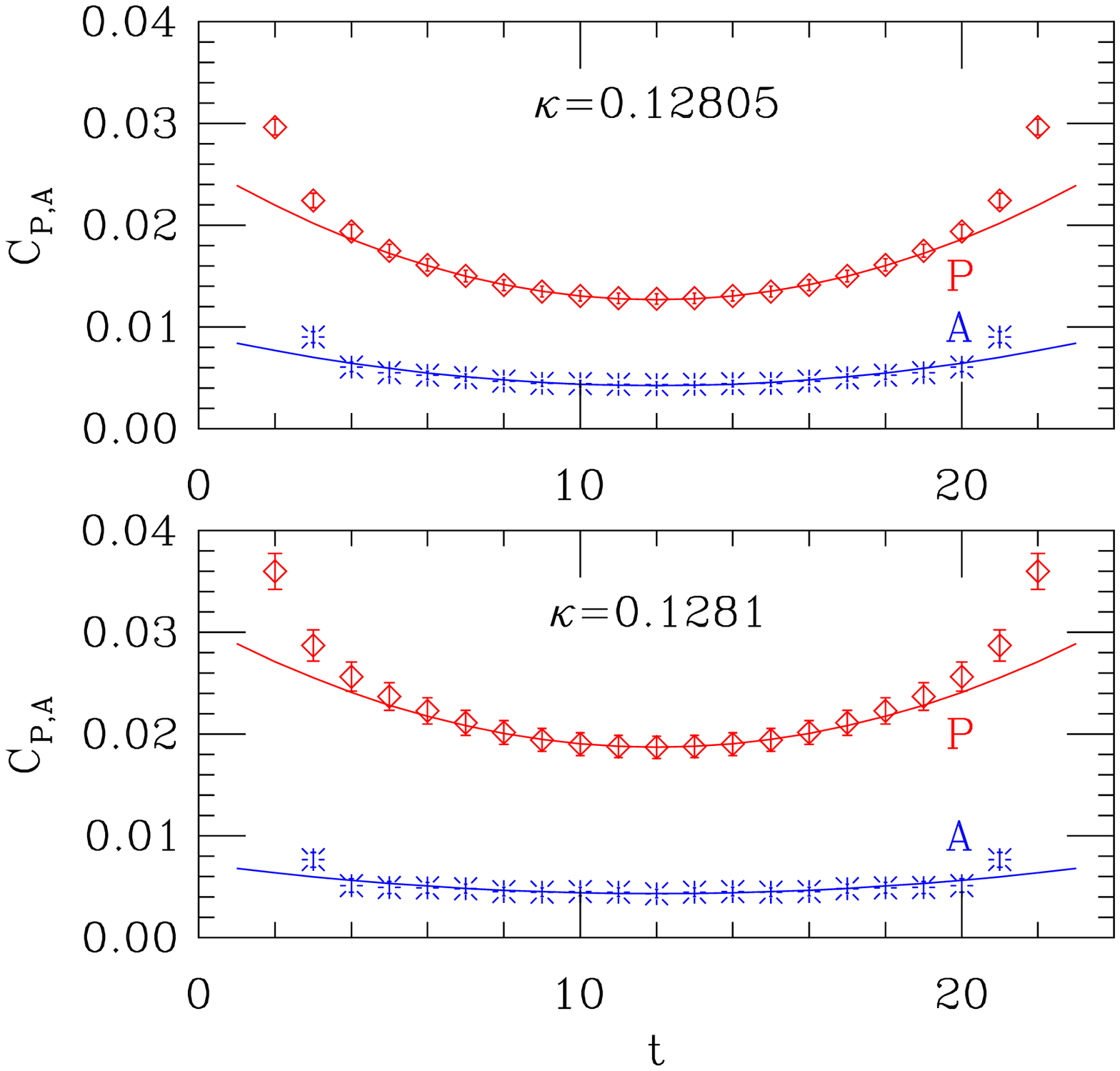}

\caption{Same as Figure \ref{cap:corr-16-1} but on the $24^{4}$ data set
at $\kappa=0.12805$ and $0.1281$.\label{cap:corr-24-1}}

\end{figure}

\begin{figure}
\includegraphics[scale=0.7]{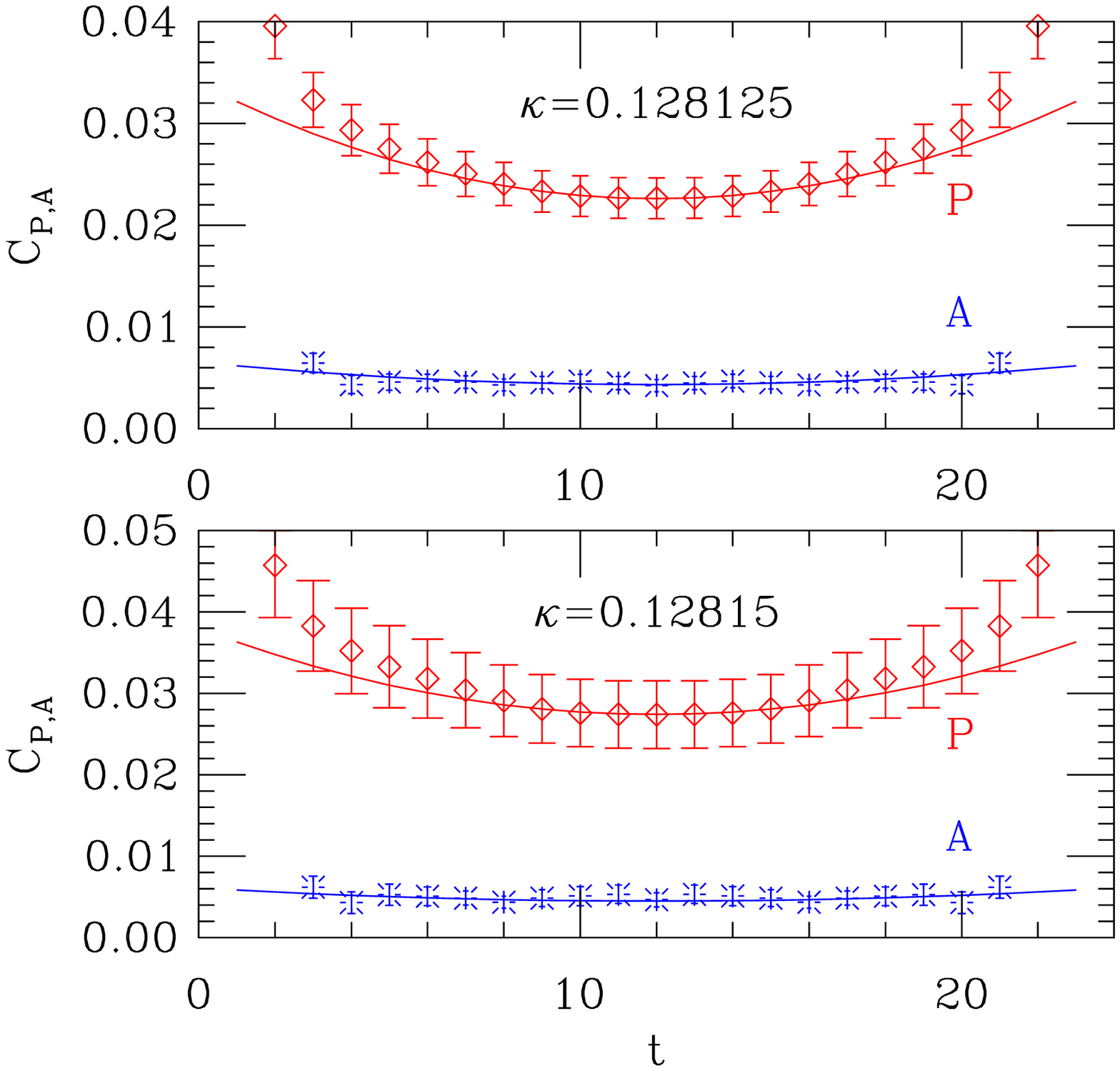}

\caption{Same as Figure \ref{cap:corr-16-1} but on the $24^{4}$ data set
at $\kappa=0.128125$ and $0.12815$.\label{cap:corr-24-2}}

\end{figure}

The results of the combined fits on the $16^{4}$ lattices are shown
in Figures \ref{cap:corr-16-1} and \ref{cap:corr-16-2}, where we
plot both the pseudoscalar and axialvector correlators (the latter
is rescaled by a factor 50 to match the scale). We use the time slices
$[5,11]$ in the fit. The data are well described by the NLO formulas
at all four mass values. The results are summarized in Table \ref{cap:Results}
where we list not only the predicted low-energy parameters but the
combination $m\Sigma V$ and an estimate for $m_{\pi}L$ as well.
We estimate the infinite volume pion mass using the GMOR relation
$m_{\pi}^{2}=N_{f}m\Sigma/F^{2}+\mathcal{O}(m^{2})$. In the $\epsilon$-regime 
one requires $m_{\pi}L\ll1$ , though according to the analysis
of Ref. \cite{Hansen:1990yg}, the $\epsilon$- and $p$-regimes
connect smoothly around $m_{\pi}L\approx2$, so values up to that
level are also acceptable. While at $\kappa=0.1278$ both $m\Sigma V$
and $m_{\pi}L$ are somewhat large, the fit indicates that the data
are described well by the $\epsilon$-regime forms at all $\kappa$
values. The predicted low-energy constants, especially F, show a slight
drift as $m$ decreases, indicating that higher order effects are
nevertheless not negligible. Considering that the expansion parameter
$1/(FL)^{2}\approx1.45$ is not at all small, this is quite possible.
The $\mathcal{O}(\epsilon^{2})$ corrections to the pseudoscalar correlator
at $t=N_{t}/2$ in Equation \ref{eq:chipt_coeff} are 34\% at $\kappa=0.1278$,
decreasing to 16\% at $\kappa=0.1281$, while the constant term $a_{a}$
of the axial correlator in Equation \ref{eq:axial_coeff} has 30-25\% corrections
in the same range. A volume of (1.85fm)$^{4}$ is not large enough
to suppress finite volume effects. %
\begin{table}
\begin{tabular}{|c|c|c|c|c|c|}
\hline 
$\kappa$  & $L$  & $m\Sigma V$  & $m_{\pi}L$ & $F${[}MeV]  & $\Sigma^{1/3}${[}MeV]\tabularnewline
\hline
\hline 
0.1278  & 16  & 3.1(2)  & 3.14 & 90(3)  & 256(6)\tabularnewline
\hline 
0.1279  & 16  & 2.1(1)  & 2.58 & 86(4)  & 254(6)\tabularnewline
\hline 
0.1280  & 16  & 1.4(1)  & 2.11 & 83(6)  & 252(7)\tabularnewline
\hline 
0.12805  & 16  & 1.0(1)  & 1.78 & 82(7)  & 250(7)\tabularnewline
\hline 
0.1281  & 16  & 0.68(5)  & 1.47 & 76(10)  & 251(7)\tabularnewline
\hline
\hline 
0.12805  & 24  & 5.2(3)  & 2.71 & 90(3)  & 248(6)\tabularnewline
\hline 
0.12810  & 24  & 3.4(2)  & 2.19 & 89(4)  & 250(6)\tabularnewline
\hline 
0.128125  & 24  & 2.6(1)  & 1.91 & 89(6)  & 248(6)\tabularnewline
\hline 
0.12815  & 24  & 2.3(1)  & 1.80 & 92(8)  & 245(8)\tabularnewline
\hline
\end{tabular}

\caption{Results from the combined fit to the pseudoscalar and axialvector
correlators. The values of $F$ and $\Sigma$ are converted to physical
units using $r_{0}=0.49$fm. The combination $m\Sigma V$ is predicted
by the fit while for  $m_{\pi}L$ we estimate the infinite volume
pion mass from the GMOR relation. \label{cap:Results}}

\end{table}

Our second data set is $24^{4},$ (2.77fm)$^{4}$, considerably larger.
The $\mathcal{O}(\epsilon^{2})$ corrections are reduced to $\approx15$\%
for the axial correlator, though the corrections are still large,
10-25\% for the pseudoscalar correlator at our mass values. Figures
\ref{cap:corr-24-1} and \ref{cap:corr-24-2} show the result of the
combined fit. Again, we find good agreement for all correlators in
the range of $5\le t\le19,$ though we use only time slices $[8,16]$
in the fit. The statistical errors are under control everywhere, though
they increase as the reweighting range increases .

The data points for $t<5$ and $t>L_{t}/a-5$ do not follow the $\epsilon$-regime
$\chi$PT predictions. A natural explanation is that the heavy excitations
that couple to the operators die out only at $t\ge5$ and influence
the correlators at small distances. If that is indeed the case, the
correlators should show similar behavior on the $16^{4}$ and $24^{4}$
sets. Indeed, at $\kappa=0.12805$ and $\kappa=0.1281$, where we
have results on both volumes, both the pseudoscalar and axialvector
correlators are identical within errors for $t<5$, showing the same
transient behavior. It is somewhat puzzling why a recent result using
overlap fermions at similar lattice spacing and even smaller quark
masses see transient behavior in the pseudoscalar channel up to $t\approx12$
\cite{Fukaya:2007pn}. It might be due to the small spatial extent
($L_{s}=16$) or the asymmetric geometry ($L_{t}=32$) used in Ref.
\cite{Fukaya:2007pn}, or that the overlap operator is more extended
and excited states die out slower.

We also have measurements of the vector current correlator. The data
and the fit quality are similar to the axialvector we presented above.
Since it does not improve the determination of the low-energy constants,
we do not include it in our analysis.

One might ask if the data, especially on the $16^{4}$
volumes, could be better fitted with NNLO, $\mathcal{O}(\epsilon^{6})$
forms. For the pseudoscalar these two-loop results are known in the
continuum \cite{Hasenfratz:1989pk}, and contain two new low-energy
constants, $L_{3}$ and $L_{4}$. On the lattice additional terms,
due to the chiral symmetry breaking of the Wilson action, also enter.
For the axialvector correlator only NLO results are available. Considering
the number of unknown parameters, it is not obvious that  a meaningful fit
could be done even if full NNLO formulas were available, though it would
be very interesting to test.

\begin{figure}
\includegraphics[scale=0.8]{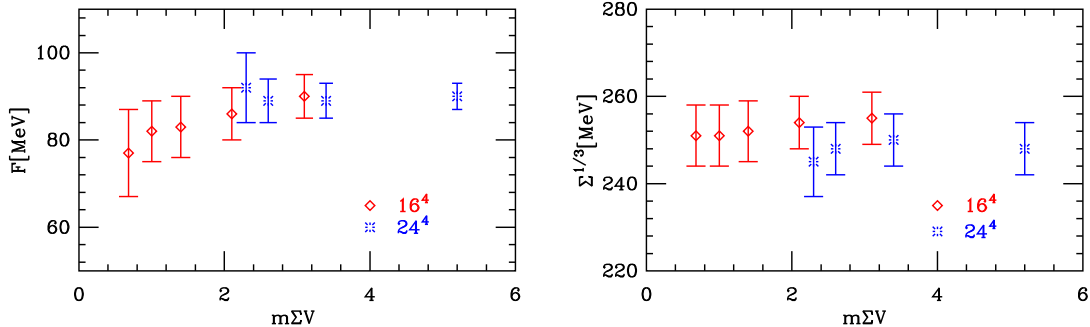}

\caption{The low-energy constant $F$ and $\Sigma^{1/3}$ as the function of
the parameter $m\Sigma V$, predicted by NLO $\chi$PT. \label{fig:Summary-plot}}

\end{figure}

We close this section with a combined plot of the low-energy constants
obtained on the two volumes, as the function of the parameter $m\Sigma V$
(Figure \ref{fig:Summary-plot}). %

As is evident both from Figure \ref{fig:Summary-plot} and Table \ref{cap:Results},
the different quark mass data on the $24^{4}$ ensemble are consistent
for both low-energy constants and the results for the chiral condensate
are consistent on the two volumes. $F$, on the other hand, shows
a drift as $m\Sigma V$ decreases. Without a large volume data point
at $m\Sigma V<2$ we cannot tell if this is due to finite volume effects,
or signals the breakdown of the $\epsilon$ expansion for $m\Sigma V>2$.
$\chi$PT formulas that connect the $\epsilon$ and $p$ regimes could
help to decide this issue. Since the next-to-leading order corrections
to $F$ are over 10\% on the $16^{4}$ data set, we prefer using the
large volume data to arrive at our final prediction, \begin{eqnarray}
F=90(4){\rm MeV,} &  & \Sigma^{1/3}=248(6){\rm MeV}\nonumber \\
Fr_{0}=0.224(10), &  & \Sigma^{1/3}r_{0}=0.617(15)\,.\label{eq:results}\end{eqnarray}
 The errors only include the statistical uncertainties.

Let us finally compare our results to other recent two flavor computations,
even though direct comparisons are problematic due to different systematic
errors. In the $p$-regime with maximally twisted mass fermions, the
ETM collaboration gets in the continuum limit $Fr_{0}=0.188(2)(7)$
and $\Sigma^{1/3}r_{0}=0.597(9)(15)$ and a compatible number for
$\Sigma$ in the $\epsilon$ regime~\cite{Dimopoulos:2007qy,Jansen:2007rx}%
\footnote{The number for $\Sigma$ is not explicitly given; the quoted errors
are obtained using $r_{0}=0.433$fm%
}. Another $\epsilon$ regime computation has been performed by JLQCD
with dynamical overlap fermions at fixed topology in a $L^{3}\times2L$,
$L=1.8{\rm fm}$ box~\cite{Fukaya:2007pn}. They get $Fr_{0}=0.217(14)$
and $\Sigma^{1/3}r_{0}=0.596(10)$. Given the statistical and systematic
errors, these results nicely agree with our determination.

Another method of extracting the low-energy constant is by looking
at the distribution of the lowest eigenvalue of the Dirac operator
and comparing it to predictions from random matrix theory. To our
knowledge, there are two such results with renormalized $N_{f}=2$
results. In Ref.~\cite{Fukaya:2007yv}, JLQCD compute $r_{0}\Sigma^{1/3}=0.624(17)(27)$
using these methods. Ref.~\cite{DeGrand:2007tm} finds $r_{0}F=0.213(11)$
and $r_{0}\Sigma^{1/3}=0.594(13)$ using nHYP link dynamical overlap
fermions. Again, there is good agreement to our findings.

\section{Conclusion}

The data presented in this paper has been generated with moderate
computer resources. This was possible due to the good chiral properties
of the action which come at relatively low cost due to the simple
nHYP smearing procedure, and the effective reweighting that allowed
us to lower the quark mass even further. Obviously, there are still
shortcomings of our analysis. The simulation is done at just one lattice
spacing, so we are not able to take our results to the continuum limit.
However, HYP smearing has improved the scaling properties of a variety
of actions. We are confident that also here cut-off effects will
be small.

Moreover, setting the scale by $r_{0}=0.49$fm is not satisfactory.
The value of $r_{0}$ is not known to high accuracy. We could use
$F$ as scale parameter. Apart from that, the $\epsilon$ regime setup
makes it problematic to use other widely used scales like the mass
of the $\Omega$ baryon.

We still are at finite lattice size and $\epsilon$-regime $\chi$PT
is a slowly converging expansion in $1/(FL)^{2}$. Here, our large
volume puts us into a good position and the comparison between the
$L/a=16$ and $L/a=24$ results shows that the finite volume effects
are under control. However, statistical errors due to the limited
number of gauge configurations are too large for a more substantiated
claim.

For the $\epsilon$ expansion to be valid, the parameter $m\Sigma V$
has to be $\mathcal{O}(1)$. Our data span the range 0.7 to 5.2 and
might go beyond the validity of the analytical expressions. An expansion
that connects the $\epsilon$ and $p$ regimes would be very useful
to control this aspect of the calculation.

Nevertheless our results are encouraging. We find that reweighting
works on a fairly large volume of $(L/a)^{4}=24^{4}$, $L\approx2.8$fm,
and the statistical fluctuations are under control despite quark masses
as low as $4$MeV. Repeating the calculation at a smaller lattice
spacing would not be prohibitively expensive and could improve on
all of the above mentioned issues.

\section{Acknowledgement}

We have benefited from discussion with O. B\"ar,
G. Colangelo, T. DeGrand, P. Hasenfratz, T. Izubuchi, S. Necco, and
F. Niedermayer. Most of the numerical work reported in this paper
was carried out at the kaon cluster at FNAL. We acknowledge the support
of the USQCD/SciDac collaboration.

The renormalization constants calculation was done based on the code
developed by T. DeGrand and Z-f. Liu. We are grateful for the permission
to use it. This research was partially supported by the US Department
of Energy and the Deutsche Forschungsgemeinschaft in the SFB/TR 09.

\bibliographystyle{apsrev} \bibliographystyle{apsrev} \bibliographystyle{apsrev}
\bibliography{lattice}

\end{document}